# Phase and transport velocities in particle and electromagnetic beams

**John Lekner**

School of Chemical and Physical Sciences, Victoria University of Wellington, PO Box 600, Wellington, New Zealand

**Abstract**
In a coherent monoenergetic beam of non-interacting particles, the phase velocity and the particle transport velocity are functions of position, with the strongest variation being in the focal region. These velocities are everywhere parallel to each other, and their product is constant in space. For a coherent monochromatic electromagnetic beam, the energy transport velocity is never greater than the speed of light, and can even be zero. The phase velocities (one each for the non-zero components of the electric and magnetic fields, in general) can be different from each other and from the energy transport velocity, both in direction and in magnitude. The phase velocities at a given point are independent of time for both particle and electromagnetic beams. The energy velocity is independent of time for the particle beam, but in general oscillates (with angular frequency $2\omega$) in magnitude and direction about its mean value at a given point in the electromagnetic beam. However, there exist electromagnetic *steady beams*, within which the energy flux, energy density and energy velocity are all independent of time.

(Some figures in this article are in colour only in the electronic version)

## 1. Introduction

A continuous beam of identical non-interacting particles of mass $m$ and energy $\hbar^2 k^2/2m$ is described by a wavefunction $\psi(\mathbf{r})e^{-i\omega t}$, where $\omega = \hbar k^2/2m$ and $\psi(\mathbf{r})$ satisfies Schrödinger's time-independent equation $(\nabla^2 + k^2)\psi = 0$. A continuous electromagnetic beam of angular frequency $\omega$ has electric and magnetic fields $\mathbf{E}(\mathbf{r}, t)$ and $\mathbf{B}(\mathbf{r}, t)$ which can be found from the complex vector potential $\mathbf{A}(\mathbf{r})e^{-i\omega t}$, each component of which satisfies the Helmholtz equation $(\nabla^2 + k^2)\psi = 0$, where $k = \omega/c$, $c$ being the speed of light [1]. (Note that in both the quantum particle beam and the electromagnetic beam cases the value of $k$ is fixed throughout the beam by the energy and angular frequency, respectively. The wavelength within the beam is not, in general, equal to $2\pi/k$ and can be very different from $2\pi/k$ within the focal region if the beam is tightly focused.)

Examples of beam wavefunctions are the approximate solution known as the Gaussian fundamental mode [2–5]:

$$\psi_G = \frac{b}{b + iz} \exp\left[ikz - \frac{k\rho^2}{2(b + iz)}\right] \quad (1)$$

where $\rho^2 = x^2 + y^2$, and a set of exact complex source/sink solutions [6–8]:

$$\psi_{\ell m} = j_\ell(kR) P_{\ell m}\left(\frac{z - ib}{R}\right) e^{\pm im\phi} \quad (2)$$

where the $j_\ell$ are spherical Bessel functions and the $P_{\ell m}$ are associated Legendre polynomials. $R$ is the distance from the complex source/sink point $(0, 0, ib) : R^2 = \rho^2 + (z - ib)^2$, and we take

$$R = (z - ib)[1 + \rho^2/(z - ib)^2]^{1/2} \quad (3)$$

in order to have $R = z - ib$ along the beam axis $\rho = 0$. Near the axis we have

$$R = z\left[1 + \frac{\rho^2}{2(z^2 + b^2)}\right] - ib\left[1 - \frac{\rho^2}{2(z^2 + b^2)}\right] + O(\rho^4) \quad (4)$$

and thus the simplest of the set (2), namely

$$\psi_{00} = j_0(kR) = \frac{\sin kR}{kR} \quad (5)$$

has (apart from the constant $kb$) the same exponent as the Gaussian $\psi_G$ of (1) on neglecting variable terms smaller by the factor $e^{-2kb}$. The same is true for the exponents of the dominant terms of all the members of the set (2); we note, however, that only those with odd $\ell - m$, for example

$$\psi_{10} = j_1(kR)P_{10}\left(\frac{z - ib}{R}\right) = \left[\frac{\sin kR}{(kR)^2} - \frac{\cos kR}{kR}\right]\frac{z - ib}{R} \quad (6)$$

have finite normalization and energy integrals in the particle and electromagnetic cases, respectively [8].

In this paper we will examine the phase and transport velocities within beams, using as examples the approximate and exact solutions (1) and (2). Please note the distinction between the usual phase and group speeds, with magnitudes $\omega/k$ and $d\omega/dk$ respectively, and the velocities defined here. For our particle beam case, $\omega = \hbar k^2/2m$, so $\omega/k = \hbar k/2m$, $d\omega/dk = \hbar k/m$, and these would be constant throughout the beam. In the electromagnetic case, with $\omega = ck$ in vacuum, both phase and group speeds would be equal to the speed of light $c$, everywhere in the beam. Instead we shall find in section 2 that

$$v_p = \frac{\hbar k^2}{2m}\frac{\nabla P}{|\nabla P|^2}, \qquad v_e = \frac{\hbar}{m}\nabla P \quad (7)$$

for the particle beam with spatial phase $P(r)$, and (in section 3) that

$$v_p = ck\frac{\nabla P}{|\nabla P|^2}, \qquad v_e = 2c\frac{E \times B}{E^2 + B^2} \quad (8)$$

for the electromagnetic beam. (Different field components may have different phases, and thus different phase velocities.) We use the subscript $e$ (for *energy*) since in the particle case $v_e$ gives the velocity of transport of particles, and hence of energy (each particle carries energy $\hbar^2 k^2/2m$), and in the electromagnetic case $v_e$ directly gives the velocity of transport of energy. The term *energy velocity* was used by Brillouin [9, section 20] in discussing the energy flow from one cell to the next in an atomic lattice. In waveguides, the power transmitted through the guide divided by the field energy per unit length gives a 'velocity of energy flow' [1, section 8.5]. This is an average over the waveguide, whereas the energy velocity used here is defined at every point within the beam. The form of the electromagnetic energy velocity as given in (8) is for Gaussian units; in SI units it would be

$$v_e = 2c^2\frac{E \times B}{E^2 + c^2B^2} \qquad \text{(SI units)}. \quad (8')$$

Some general results follow immediately from (7) and (8). In the particle beam case, the phase and energy (particle transport) velocities are everywhere parallel to each other, and their product is equal to the constant (particle energy /particle mass):

$$v_p \cdot v_e = v_p v_e = \frac{\hbar^2 k^2}{2m^2}. \quad (9)$$

For electromagnetic beams, the energy transport velocity cannot exceed the speed of light, as expected:

$$\frac{v_e^2}{c^2} = \frac{4E^2B^2 - 4(E \cdot B)^2}{(E^2 + B^2)^2}$$

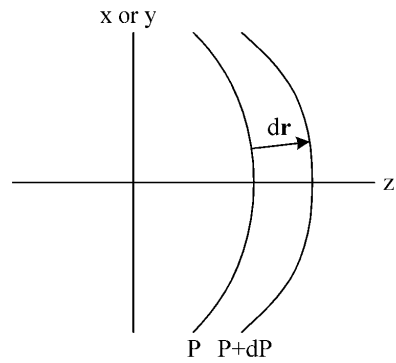

**Figure 1.** Phase velocity: the total phase $P(r) - \omega t$ is constant on a given wavefront, so $dP = \omega \, dt$. But $dP = |\nabla P|dr$, so the phase velocity has magnitude $dr/dt = \omega/|\nabla P|$ and the direction of $\nabla P$.

so

$$1 - \frac{v_e^2}{c^2} = \frac{(E^2 - B^2)^2 + 4(E \cdot B)^2}{(E^2 + B^2)^2} \geqslant 0. \quad (10)$$

Note that the phase function $P$ is a function of space, not of time: we write the wavefunction(s) of the beam as $\psi(r)e^{-i\omega t}$, with modulus $M$ and phase $P$:

$$\psi(r) = M(r)e^{iP(r)}. \quad (11)$$

Thus both the phase and particle transport velocities are independent of time in the particle beam, while only the phase velocities are independent of time in the electromagnetic case, in general. There exist *steady beams*, to be discussed in section 4, for which the energy density, Poynting vector and energy velocity are all independent of time, but these are a special set. In general, all of these three quantities oscillate in time about their mean values.

## 2. Phase and particle transport velocities, scalar beams

We consider the phase velocity first. The phase $P(r)$ has equiphase surfaces fixed in space (see figure 1); the total phase of the wavefunction is $P(r) - \omega t$. In time $dt$ the wavefront moves from the $P$ to the $P + dP$ surfaces, in the direction of $\nabla P$ so the $dP = |\nabla P|dr = \omega \, dt$, since the total phase function $P - \omega t$ is fixed for a given wavefront.

Hence the phase speed is $\omega/|\nabla P|$ and the phase velocity is

$$v_p = \frac{\omega \nabla P}{|\nabla P|^2} = \frac{\hbar k^2}{2m}\frac{\nabla P}{|\nabla P|^2}. \quad (12)$$

The above arguments are similar to those given in section 1.3.3 of [10], with the difference that we associate a direction (that of $\nabla P$) with our $v_p$, whereas Born and Wolf emphatically do not with theirs, which is $\omega/|\nabla P|$: 'the phase velocity does not behave as a vector' [10, p 19].

The particle transport velocity can be defined in terms of the probability current density (see, for example, [11, section 3.1])

$$J = \frac{\hbar}{m}\text{Im}(\psi^*\nabla\psi) = \frac{\hbar}{m}M^2\nabla P. \quad (13)$$

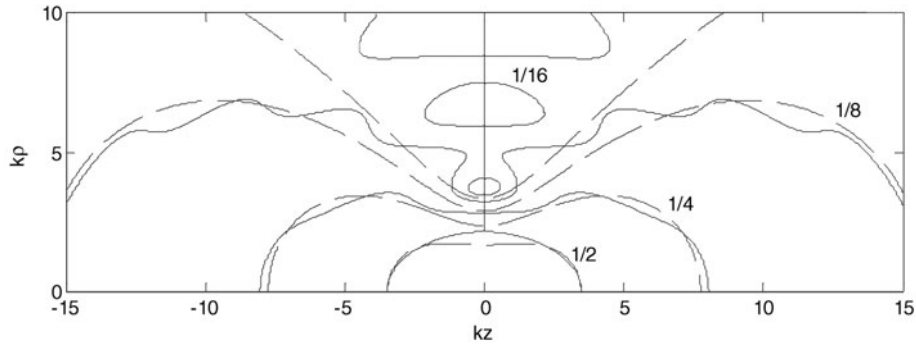

**Figure 2.** Surfaces of constant modulus for $\psi_G$ (- - -), and $(\beta/\sinh\beta)\psi_{00}$, (———) drawn $\beta = kb = 2$. Both wavefunctions are normalized to unity at the origin. Note the zeros of $\psi_{00}$ in the focal plane $z = 0$, at $k\sqrt{\rho^2 - b^2} = n\pi$, i.e. at $k\rho = \sqrt{\beta^2 + n^2\pi^2}$. The three-dimensional surfaces are obtained by rotating the diagram about the $z$ axis.

We interpret $\boldsymbol{J}$ as $|\psi|^2$ times a velocity of particle or energy transport, $\boldsymbol{v}_e$, i.e. $\boldsymbol{J} = M^2 \boldsymbol{v}_e$, so

$$\boldsymbol{v}_e = \frac{\hbar}{m}\nabla P. \tag{14}$$

By way of example, consider the beams represented by $\psi_G$ and $\psi_{00}$ given in equations (1) and (2). We rewrite these wavefunctions in terms of modulus and phase:

$$\psi_G = \frac{b}{\sqrt{b^2 + z^2}} \exp\left[\frac{-kb\rho^2}{2(b^2 + z^2)}\right]$$
$$\times \exp i\left\{kz - \mathrm{atn}\left(\frac{z}{b}\right) + \frac{kz\rho^2}{2(b^2 + z^2)}\right\} \tag{15}$$

$$\psi_{00} = \frac{(\sin^2\beta\xi + \sinh^2\beta\eta)^{1/2}}{\beta(\xi^2 + \eta^2)^{1/2}}$$
$$\times \exp i\left\{\beta\xi - \mathrm{atn}\frac{\xi}{\eta} + \mathrm{atn}\left[\frac{\sin 2\beta\xi}{\mathrm{e}^{2\beta\eta} - \cos 2\beta\xi}\right]\right\}. \tag{16}$$

The $\psi_{\ell m}$ are expressed most simply in terms of oblate spheroidal coordinates $\xi$ and $\eta$ [7, 8], since these are respectively proportional to the real and imaginary parts of the complex distance $R$ given in (3):

$$R = (\xi - i\eta)b, \qquad \rho^2 = (1+\xi^2)(1-\eta^2)b^2, \qquad z = \xi\eta b. \tag{17}$$

The inverse relations are, with $s^2 = \rho^2 + z^2 - b^2$,

$$2b^2\xi^2 = [s^4 + 4b^2z^2]^{1/2} + s^2, \qquad 2b^2\eta^2 = [s^4 + 4b^2z^2]^{1/2} - s^2. \tag{18}$$

On the beam axis $\rho = 0$ we have $\eta = 1$ and $\xi = z/b$. (Here we take $-\infty < \xi < \infty$, $0 \leqslant \eta \leqslant 1$, rather than the alternative choice $0 \leqslant \xi < \infty$, $-1 \leqslant \eta \leqslant 1$ [12].) Finally, in (16) we have set the dimensionless parameter $kb$ equal to $\beta$.

The parameter $\beta$ determines the divergence half-angle of the beam: when $b^2$ and $\rho^2$ are much smaller than $z^2$ the exponent in the modulus of $\psi_G$ tends to $-\beta\rho^2/2z^2$, so the beam amplitude falls to $e^{-1}$ from its axial value at $\rho^2 = 2z^2/\beta$, from which we see that the beam divergence half-angle is $\theta = \mathrm{atn}(\frac{2}{\beta})^{1/2}$. For the $\psi_{\ell m}$ beams, the exponent in the modulus tends to the same function when $\beta$ is large, so the same divergence angle applies. For $\beta$ small compared to unity, the oscillatory term $\sin\beta\xi = \sin[kz + \beta\rho^2/[2(z^2 + b^2)] + O(\rho^4)]$ becomes as important as the hyperbolic term $\sinh\beta\eta =$ $\sinh\left[\beta - \frac{\beta\rho^2}{2(z^2 + b^2)} + O(\rho^4)\right]$, and the amplitude decay ceases to be predominantly exponential in $\rho^2$.

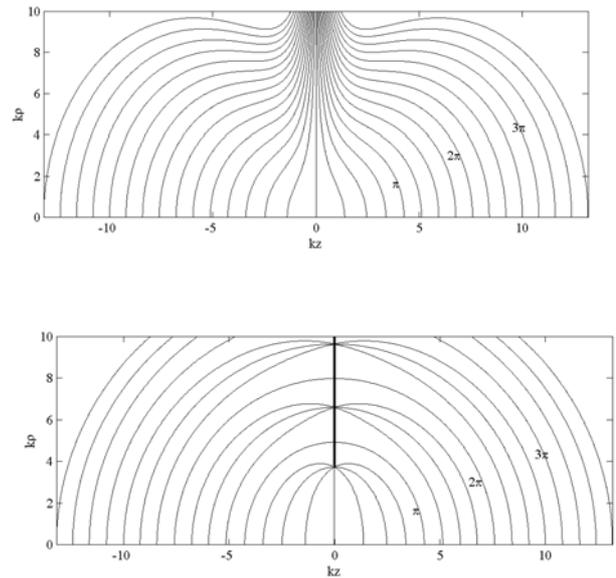

**Figure 3.** Surfaces of constant phase for $\psi_G$ (upper figure) and $\psi_{00}$ (lower figure), drawn for $\beta = 2$ at increments of $\pi/4$. The equiphase surfaces for $\psi_G$ all go off to infinite $\rho$ in the $z = 0$ plane, while those of $\psi_{00}$ converge onto the zeros of $\psi_{00}$, which lie on the circles $\rho_n = \sqrt{b^2 + (n\pi/k)^2} = b\sqrt{1 + (n\pi/\beta)^2}$. The surfaces with phase $P$ between 0 and $\pi$ converge onto $\rho_1$, those with $P$ between $\pi$ and $2\pi$ onto $\rho_2$, etc. The surfaces with phase equal to an integer multiple of $\pi$ converge onto the circles $\rho = b\sqrt{1 + (X/\beta)^2}$, where $\tan X = X$.

Figure 2 shows surfaces of constant modulus for $\psi_G$ and $\psi_{00}$, drawn for $kb = \beta = 2$ (i.e. for a beam divergence half-angle of 45°). Note the zeros of $\psi_{00}$ in the focal plane, on the circles $k\sqrt{\rho^2 - b^2} = n\pi$, $n$ a positive integer. The equiphase surfaces compared in figure 3 correspondingly converge (for $\psi_{00}$) onto these circles of zero modulus, where the phase is undefined. In contrast, the constant-phase surfaces of $\psi_G$ go off to infinity.

Figure 4 compares the probability current densities for $\psi_G$ and $(\beta/\sinh\beta)\psi_{00}$, the factor $\beta/\sinh\beta$ being inserted so that both wavefunctions are normalized to unity at the origin (which is the centre of the focal plane). In both cases the current

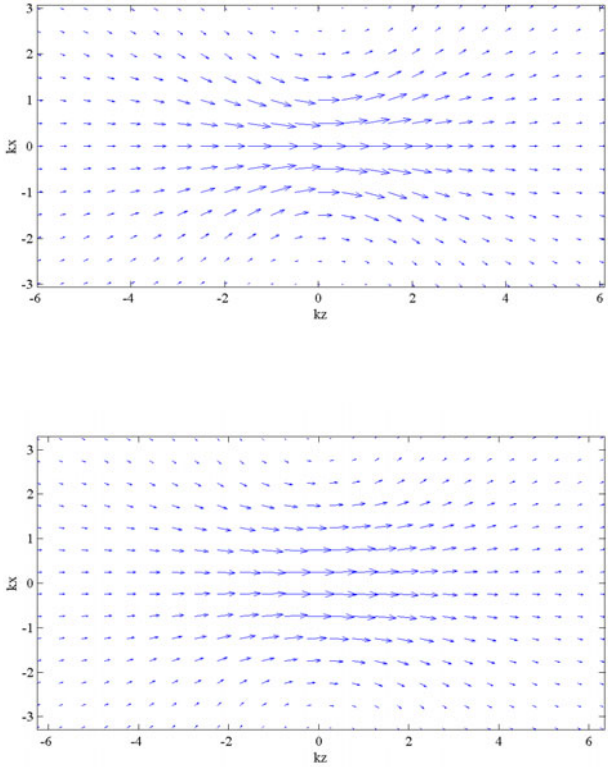

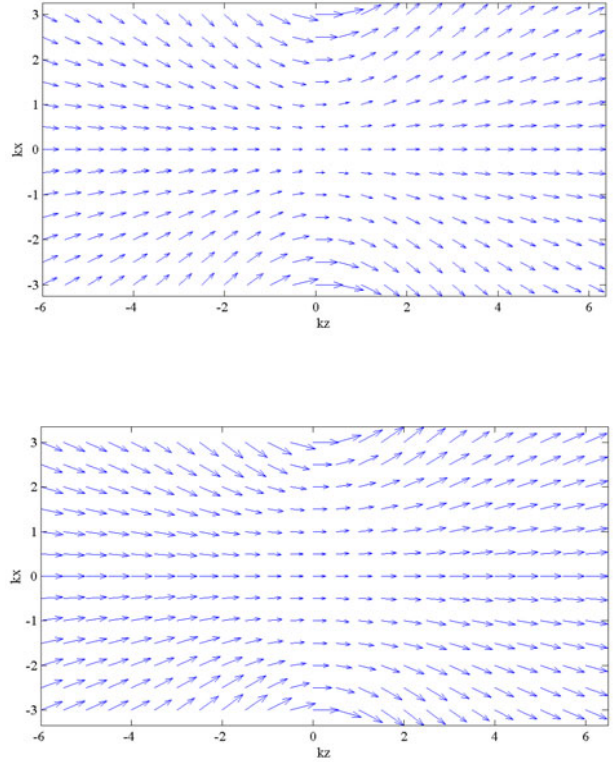

**Figure 4.** Probability current densities for $\psi_G$ (upper figure) and $(\beta/\sinh\beta)\psi_{00}$ (lower figure), drawn for $\beta = 2$.

**Figure 5.** Energy velocity fields for $\psi_G$ (upper figure) and $\psi_{00}$ (lower figure), drawn for $\beta = 2$.

density is maximum at the origin, with respective values

$$J_G(0,0) = \frac{\hbar k}{m}\left(1 - \frac{1}{\beta}\right),$$
$$J_{00}(0,0) = \frac{\hbar k}{m}\frac{e^{2\beta}(\beta - 1) + \beta + 1}{\beta(e^{2\beta} - 1)}. \quad (19)$$

We note that the approximate Gaussian wavefunction fails for small $\beta(= kb)$: the current would go negative at the origin for $\beta < 1$ and diverge to $-\infty$ as $\beta \to 0$. The $\psi_{00}$ current is well behaved as $\beta \to 0$:

$$J_{00}(0,0) = \frac{\hbar k}{m}\left[\frac{1}{3}\beta - \frac{1}{45}\beta^3 + O(\beta^5)\right]. \quad (20)$$

(The probability current density goes to zero with $\beta$ since $\psi_{00}$ for small $\beta$ represents almost equal amounts of forward and backward propagation.) At large $\beta$ the currents at the origin both tend to $\hbar k/m$, as one would expect from a broad beam normalized to unity at the centre of its focal plane. Because both $\psi_G$ and $(\beta/\sinh\beta)\psi_{00}$ have unit modulus at the origin, the energy velocity $v_e$ at the origin has the magnitude given in (19) for the current $J$. The phase velocities at the origin have magnitudes $v_p = (\hbar k/2m)(\hbar k/m v_e)$, i.e.

$$v_G^{(p)}(0,0) = \frac{\hbar k}{2m}\frac{\beta}{\beta - 1},$$
$$v_{00}^{(p)}(0,0) = \frac{\hbar k}{2m}\frac{\beta(e^{2\beta} - 1)}{e^{2\beta}(\beta - 1) + \beta + 1}. \quad (21)$$

Again we see the failure of the approximate Gaussian beam wavefunction, with divergence in the corresponding phase velocity at $\beta = 1$. The $\psi_{00}$ beam has $v_p \to \frac{3}{2}\hbar/mb$ for $\beta \to 0$ and $v_p \to \hbar k/2m$ for large $\beta$. Figures 5 and 6 show the energy and phase velocity fields of $\psi_G$ and $\psi_{00}$.

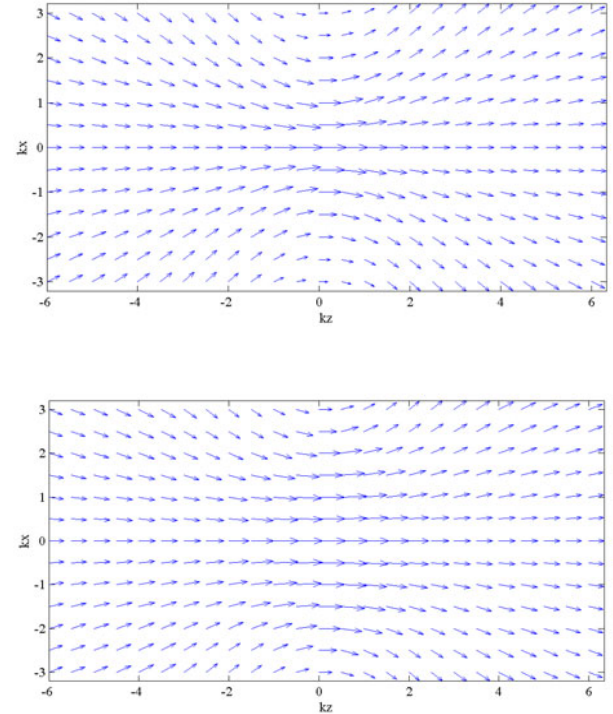

**Figure 6.** Phase velocity fields of $\psi_G$ (upper figure) and $\psi_{00}$ (lower figure), drawn for $\beta = 2$.

The results for the beam wavefunction $\psi_{10}$ given in (6) are similar for large $\beta$ but different at small $\beta$: the probability current density of a beam normalized to unity at the origin (i.e. with wavefunction $\psi_{10}/j_1(-i\beta)$) is

$$J_{10}(0,0) = \frac{\hbar k}{m} \frac{\beta^3 \cosh\beta \sinh\beta - (\beta^2 - 2)\cosh^2\beta - (\beta^2 + 2)}{\beta[(\beta^2 - 1)\cosh^2\beta + 1]}. \quad (22)$$

This tends to $\hbar/mb$ as $\beta$ tends to zero and to $\hbar k/m$ for large $\beta$. The energy velocity at the origin is also given by (22), since the modulus is unity there. The phase velocity at the origin is, from (9) or (12), $\hbar k/2m$ times the reciprocal of the function multiplying $\hbar k/m$ in (22). The phase velocity thus tends to $\hbar k^2 b/2m$ for small $\beta$ and to $\hbar k/2m$ for large $\beta$.

## 3. Phase and energy velocities for electromagnetic beams

In the Lorentz gauge, and with all time dependence in the factor $e^{-i\omega t}$, the complex electric and magnetic fields can be obtained in terms of spatial derivatives of the complex vector potential [1, 8]

$$\boldsymbol{B}(\boldsymbol{r}) = \nabla \times \boldsymbol{A}(\boldsymbol{r}), \qquad \boldsymbol{E}(\boldsymbol{r}) = \frac{i}{k}[\nabla(\nabla \cdot \boldsymbol{A}(\boldsymbol{r})) + k^2 \boldsymbol{A}(\boldsymbol{r})]. \quad (23)$$

The real fields are, for example,

$$\boldsymbol{E}(\boldsymbol{r}, t) = \text{Re}\{\boldsymbol{E}(\boldsymbol{r})e^{-i\omega t}\} = \tfrac{1}{2}\lfloor \boldsymbol{E}(\boldsymbol{r})e^{-i\omega t} + \boldsymbol{E}^*(\boldsymbol{r})e^{i\omega t} \rfloor. \quad (24)$$

In terms of the real fields, the energy density and Poynting vector (which gives the energy flow through unit area in unit time) are given by

$$u = \frac{1}{8\pi}(E^2 + B^2), \qquad \boldsymbol{S} = \frac{c}{4\pi}\boldsymbol{E} \times \boldsymbol{B}. \quad (25)$$

The corresponding expressions in terms of the complex fields, averaged over one cycle, are

$$\overline{u} = \frac{1}{16\pi}[\boldsymbol{E}(\boldsymbol{r}) \cdot \boldsymbol{E}^*(\boldsymbol{r}) + \boldsymbol{B}(\boldsymbol{r}) \cdot \boldsymbol{B}^*(\boldsymbol{r})] \quad (26)$$

$$\overline{\boldsymbol{S}} = \frac{c}{16\pi}[\boldsymbol{E}(\boldsymbol{r}) \times \boldsymbol{B}^*(\boldsymbol{r}) + \boldsymbol{E}^*(\boldsymbol{r}) \times \boldsymbol{B}(\boldsymbol{r})]. \quad (27)$$

Each component of $\boldsymbol{A}$ satisfies the Helmholtz equation $(\nabla^2 + k^2)\psi = 0$, with $k = \omega/c$. There are many possibilities for beams, the simplest being $\boldsymbol{A} = (0, 0, A_0\psi)$, which gives the transverse-magnetic (TM) beam. When $\psi$ is cylindrically symmetric, as it is for our three examples $\psi_G$, $\psi_{00}$ and $\psi_{10}$, the corresponding TM fields are

$$\boldsymbol{B} = A_0 \frac{\partial \psi}{\partial \rho}(\sin\phi, -\cos\phi, 0) \quad (28)$$

$$\boldsymbol{E} = \frac{iA_0}{k}\left(\cos\phi \frac{\partial^2\psi}{\partial\rho\,\partial z}, \sin\phi \frac{\partial^2\psi}{\partial\rho\,\partial z}, \frac{\partial^2\psi}{\partial z^2} + k^2\psi\right). \quad (29)$$

In general, each component of $\boldsymbol{B}$ and $\boldsymbol{E}$ will have its own phase (and thus its own phase velocity) when the complex field component is written as a modulus times a phase factor, but in the TM example above $B_x$ and $B_y$ share a common phase function, as do $E_x$ and $E_y$. Thus there are three wavefronts associated with a TM beam with cylindrical symmetry: those of $\{B_x, B_y\}$, $\{E_x, E_y\}$ and $E_z$. For a given phase function $P(\boldsymbol{r})$, the phase velocity is

$$\boldsymbol{v}_p = \frac{\omega \nabla P}{|\nabla P|^2} = c\frac{k\nabla P}{|\nabla P|^2} \quad (30)$$

by the arguments given in section 2.

There is only one energy velocity field: the energy flux is given by the Poynting vector $\boldsymbol{S}$, and in free space this is related to the energy density $u$ by the conservation law (see, for example, section 6.8 of [1])

$$\frac{\partial u}{\partial t} + \nabla \cdot \boldsymbol{S} = 0. \quad (31)$$

We define the energy velocity by analogy with fluid dynamics, in which mass conservation is $\frac{\partial \rho}{\partial t} + \nabla \cdot (\rho \boldsymbol{v}) = 0$, where $\rho$ is the mass density and $\boldsymbol{v}$ is the velocity of fluid flow. Thus

$$\boldsymbol{v}_e = \frac{\boldsymbol{S}}{u} = 2c\frac{\boldsymbol{E} \times \boldsymbol{B}}{E^2 + B^2}. \quad (32)$$

In section 6.8 of [1] it is stated 'since only its divergence appears in the conservation law, the Poynting vector is arbitrary to the extent that the curl of any vector field can be added to it. Such an added term can, however, have no physical consequences'. We have taken the customary choices for energy density and energy flux, and have seen (in section 1) that these choices lead to the satisfactory result that the energy velocity in an electromagnetic field cannot exceed $c$. The proof of this result, in equations (10), would, however, fail if we added the curl of a vector field to $\boldsymbol{E} \times \boldsymbol{B}$. We conclude that we must of necessity omit such a term to retain the relativistic requirement of $v_e \leqslant c$.

Both the energy flux and the energy density oscillate about their mean values $\overline{\boldsymbol{S}}$ and $\overline{u}$ at angular frequency $2\omega$, except in *steady beams*, to be discussed in the next section. Thus $\boldsymbol{v}_e$ will likewise oscillate about its mean value $\overline{\boldsymbol{v}}_e$, in general. At each point in space and time the magnitude of $\boldsymbol{v}_e$ will not exceed the speed of light, as we saw in section 1.

## 4. Steady beams

In all electromagnetic waves the fields $\boldsymbol{E}$ and $\boldsymbol{B}$ must oscillate in time. There are, however, monochromatic beams, which we shall call *steady beams*, in which $\boldsymbol{E} \times \boldsymbol{B}$ and $E^2 + B^2$ are everywhere independent of time. A particular case was noted in section 4 of [8]; here we shall generalize this idea. Let the complex vector potential $\boldsymbol{A}(\boldsymbol{r})$ lead to the complex fields $\boldsymbol{E}(\boldsymbol{r})$ and $\boldsymbol{B}(\boldsymbol{r})$ via (23). Then the dual potential (all components of which also satisfy the Helmholtz equation)

$$\boldsymbol{A}' = (ik)^{-1}\nabla \times \boldsymbol{A} = (ik)^{-1}\boldsymbol{B} \quad (33)$$

leads to the fields

$$\boldsymbol{B}' = \nabla \times \boldsymbol{A}' = (ik)^{-1}\nabla \times \boldsymbol{B} = -\boldsymbol{E}$$
$$\boldsymbol{E}' = \frac{i}{k}[\nabla(\nabla \cdot \boldsymbol{A}') + k^2 \boldsymbol{A}'] \quad (34)$$
$$= \left(\frac{i}{k}\right)\left(\frac{1}{ik}\right)[\nabla(\nabla \cdot \boldsymbol{B}) + k^2 \boldsymbol{B}] = \boldsymbol{B}$$

495

where we have used the source-free Maxwell equations $\nabla \times \boldsymbol{B} - \frac{1}{c}\partial \boldsymbol{E}/\partial t = 0$ and $\nabla \cdot \boldsymbol{B} = 0$ in the first and second parts of (34), respectively. This is the simple duality transformation $\boldsymbol{E} \to \boldsymbol{B}, \boldsymbol{B} \to -\boldsymbol{E}$ (for the general transformation, see section 6.12 of [1]) under which the Maxwell equations are invariant. Now consider the vector potential $\boldsymbol{A}'' = \boldsymbol{A} + \mathrm{i}\boldsymbol{A}'$. This gives the fields $\boldsymbol{E}'' = \boldsymbol{E} + \mathrm{i}\boldsymbol{B}, \boldsymbol{B}'' = \boldsymbol{B} - \mathrm{i}\boldsymbol{E}$, so that

$$\boldsymbol{E}'' = \mathrm{i}\boldsymbol{B}''. \tag{35}$$

Likewise the combination $\boldsymbol{A} - \mathrm{i}\boldsymbol{A}'$ leads to $\boldsymbol{E}'' = -\mathrm{i}\boldsymbol{B}''$. Fields for which (35) is true have (we drop the double primes)

$$\begin{aligned}\boldsymbol{B}(\boldsymbol{r},t) &= \mathrm{Re}\{\boldsymbol{B}(\boldsymbol{r})\mathrm{e}^{-\mathrm{i}\omega t}\} \\ &= \mathrm{Re}\{(\boldsymbol{B}_r + \mathrm{i}\boldsymbol{B}_i)(\cos\omega t - \mathrm{i}\sin\omega t)\} \\ &= \boldsymbol{B}_r\cos\omega t + \boldsymbol{B}_i\sin\omega t \\ \boldsymbol{E}(\boldsymbol{r},t) &= \mathrm{Re}\{\mathrm{i}\boldsymbol{B}(\boldsymbol{r})\mathrm{e}^{-\mathrm{i}\omega t}\} \\ &= \mathrm{Re}\{(\mathrm{i}\boldsymbol{B}_r - \boldsymbol{B}_i)(\cos\omega t - \mathrm{i}\sin\omega t)\} \\ &= \boldsymbol{B}_r\sin\omega t - \boldsymbol{B}_i\cos\omega t\end{aligned} \tag{36}$$

where $\boldsymbol{B}_r(\boldsymbol{r})$ and $\boldsymbol{B}_i(\boldsymbol{r})$ are the real and imaginary parts of the complex field $\boldsymbol{B}(\boldsymbol{r})$, and $\boldsymbol{E}_r = -\boldsymbol{B}_i, \boldsymbol{E}_i = \boldsymbol{B}_r$. The resulting energy density and flux are time-independent:

$$u = \frac{1}{8\pi}(B_r^2 + B_i^2), \qquad \boldsymbol{S} = \frac{c}{4\pi}\boldsymbol{B}_r \times \boldsymbol{B}_i \tag{37}$$

(when $\boldsymbol{E} = -\mathrm{i}\boldsymbol{B}, \boldsymbol{S}$ becomes $(c/4\pi)\boldsymbol{B}_i \times \boldsymbol{B}_r$).

The 'steady beams' thus have $\boldsymbol{A} \pm k^{-1}\nabla \times \boldsymbol{A}$ as a vector potential (where each component of $\boldsymbol{A}$ must satisfy the Helmholtz equation), electric and magnetic fields which are in phase quadrature and equal in magnitude ($\boldsymbol{E} = \pm\mathrm{i}\boldsymbol{B}$), and time-independent energy density and flux.

The relations $\boldsymbol{E} = \pm\mathrm{i}\boldsymbol{B}$ are necessary as well as sufficient for the time independence of the energy flux and energy density:

$$\begin{aligned}\frac{4\pi}{c}\boldsymbol{S} &= \boldsymbol{E} \times \boldsymbol{B} \\ &= (\boldsymbol{E}_r\cos\omega t + \boldsymbol{E}_i\sin\omega t)(\boldsymbol{B}_r\cos\omega t + \boldsymbol{B}_i\sin\omega t) \\ &= \boldsymbol{E}_r \times \boldsymbol{B}_r\cos^2\omega t + (\boldsymbol{E}_r \times \boldsymbol{B}_i + \boldsymbol{E}_i \times \boldsymbol{B}_r)\cos\omega t\sin\omega t \\ &\quad + \boldsymbol{E}_i \times \boldsymbol{B}_i\sin^2\omega t\end{aligned} \tag{38}$$

$$\begin{aligned}8\pi u &= \boldsymbol{E}^2 + \boldsymbol{B}^2 \\ &= (\boldsymbol{E}_r\cos\omega t + \boldsymbol{E}_i\sin\omega t)^2 + (\boldsymbol{B}_r\cos\omega t + \boldsymbol{B}_i\sin\omega t)^2 \\ &= (E_r^2 + B_r^2)\cos^2\omega t + 2(\boldsymbol{E}_r \cdot \boldsymbol{E}_i + \boldsymbol{B}_r \cdot \boldsymbol{B}_i)\cos\omega t\sin\omega t \\ &\quad + (E_i^2 + B_i^2)\sin^2\omega t\end{aligned} \tag{39}$$

and the eight equations $\boldsymbol{E}_r \times \boldsymbol{B}_r = \boldsymbol{E}_i \times \boldsymbol{B}_i, \boldsymbol{E}_r \times \boldsymbol{B}_i + \boldsymbol{E}_i \times \boldsymbol{B}_r = 0, E_r^2 + B_r^2 = E_i^2 + B_i^2, \boldsymbol{E}_r \cdot \boldsymbol{E}_i + \boldsymbol{B}_r \cdot \boldsymbol{B}_i = 0$ to be satisfied among the six components of $\boldsymbol{E}_r, \boldsymbol{E}_i$ ($\boldsymbol{B}_r, \boldsymbol{B}_i$ having been specified) are solved by $\{\boldsymbol{E}_r = \mp\boldsymbol{B}_i, \boldsymbol{E}_i = \pm\boldsymbol{B}_r\}$ and by no other real set. (The solutions $\boldsymbol{E}_r = \pm\mathrm{i}\boldsymbol{B}_r, \boldsymbol{E}_i = \pm\mathrm{i}\boldsymbol{B}_i$ do not apply.)

We note that, when $\boldsymbol{E} = \pm\mathrm{i}\boldsymbol{B}$, both $\boldsymbol{E}$ and $\boldsymbol{B}$ are eigenstates of curl, with eigenvalue $\pm k$:

$$\nabla \times \boldsymbol{B} = \pm k\boldsymbol{B} \qquad \nabla \times \boldsymbol{E} = \pm k\boldsymbol{E} \tag{40}$$

(or $\nabla \times \boldsymbol{B}_r = \pm k\boldsymbol{B}_r, \nabla \times \boldsymbol{B}_i = \pm k\boldsymbol{B}_i$, etc). These relations follow by substituting $\boldsymbol{E} = \pm\mathrm{i}\boldsymbol{B}$ or equivalently $\{\boldsymbol{E}_r = \mp\boldsymbol{B}_i, \boldsymbol{E}_i = \pm\boldsymbol{B}_r\}$ into the Maxwell curl equations.

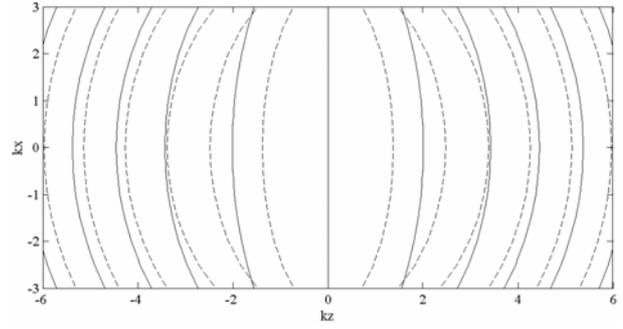

**Figure 7.** Comparison of the equiphase surfaces $P = $ constant (- - -) and $P_E = $ constant (——) for the TE beam with $\psi_{00} = \sin(kR)/kR$, drawn for $\beta = kb = 2$ at increments of $\pi/4$. Both phases are zero in the focal plane $z = 0$. Note the marked difference in phase value and in curvature when $\beta = 2$. For $\beta \gg 1$ and $z^2 \ll b^2$, $P_E \approx P$, on the beam axis.

## 5. TE beam phases, wavefront curvatures and phase velocities

We shall now calculate the phases, phase velocities and energy velocities of the azimuthally symmetric TE and 'TEM' beams, beginning with the transverse electric beam. This has electric field lines which are circles concentric with the beam axis [8, section 4]:

$$\boldsymbol{E} = A_0\frac{\partial\psi}{\partial\rho}(\sin\phi, -\cos\phi, 0). \tag{41}$$

The magnetic field is the sum of a radial vector in the $xy$ plane plus a longitudinal component:

$$\boldsymbol{B} = \frac{A_0}{\mathrm{i}k}\left(\cos\phi\frac{\partial^2\psi}{\partial\rho\,\partial z}, \sin\phi\frac{\partial^2\psi}{\partial\rho\,\partial z}, \frac{\partial^2\psi}{\partial z^2} + k^2\psi\right). \tag{42}$$

Thus $\boldsymbol{E}$ has one family of equiphase surfaces associated with it, $\boldsymbol{B}$ has two (one for the two transverse components, another for the longitudinal component). Let us consider the phase of the electric vector: from (40), with $\psi = M\mathrm{e}^{\mathrm{i}P}$ and assuming a real $A_0$,

$$\mathrm{ph}(\boldsymbol{E}) = \mathrm{ph}(\partial\psi/\partial\rho) = \mathrm{ph}\left\{\left(\frac{\partial M}{\partial\rho} + \mathrm{i}M\frac{\partial P}{\partial\rho}\right)\mathrm{e}^{\mathrm{i}P}\right\} \tag{43}$$

or

$$P_E = P + \mathrm{atn}\left\{\frac{M\partial P/\partial\rho}{\partial M/\partial\rho}\right\}. \tag{44}$$

Figure 7 shows equiphase surfaces $P = $ constant and $P_E = $ constant for $\psi_{00} = j_0(kR)$. In terms of the oblate spheroidal coordinates $\xi, \eta$ of (17) and (18) we have

$$P = \beta\xi + \mathrm{atn}\left\{\frac{\sin 2\beta\xi}{\mathrm{e}^{2\beta\eta} - \cos 2\beta\xi}\right\} - \mathrm{atn}\left(\frac{\xi}{\eta}\right) \tag{45}$$

and

$$\begin{aligned}P_E = P &- \mathrm{atn}\{\{\beta\xi(\xi^2+\eta^2)CS - 2\xi\eta(C^2-c^2) \\ &+ \beta\eta(\xi^2+\eta^2)cs\}\{\beta\eta(\xi^2+\eta^2)CS \\ &+ (\xi^2-\eta^2)(C^2-c^2) - \beta\xi(\xi^2+\eta^2)cs\}^{-1}\}\end{aligned} \tag{46}$$

where $C = \cosh\beta\eta, S = \sinh\beta\eta, c = \cos\beta\xi, s = \sin\beta\xi$.

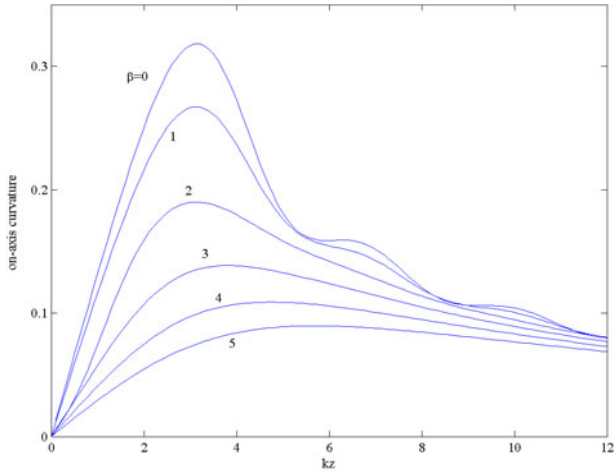

**Figure 8.** On-axis curvature $\kappa_0(z)/k$ of the phase surfaces of $\psi_{00}$, drawn for $\beta = 0$–5. The limiting form as $\beta \to 0$ is given by the last line of equation (51).

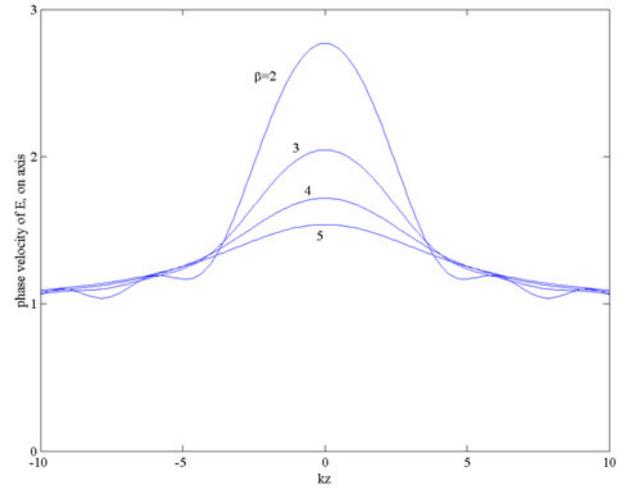

**Figure 9.** On-axis phase velocity of the transverse electric field components of the TE beam, drawn for $\beta = 2$–5. The curves show $v_p^E(z)/c$; note that the phase velocity can be substantially larger than the speed of light in the focal region.

The curvature of the equiphase surfaces is important in the design of laser resonators [13, 14]. On a given cylindrically symmetric equiphase surface $P = $ constant, the distance from the axis $\rho$ is a known function of $z$, and the curvature of the surface is given by the well-known formula

$$\kappa(\rho, z) = \frac{d^2\rho/dz^2}{[1 + (d\rho/dz)^2]^{3/2}}. \quad (47)$$

The derivatives of $\rho$ with respect to $z$ can be expressed in terms of partial derivatives of the phase function (compare section 4 of [15]):

$$\frac{d\rho}{dz} = -\frac{\partial P/\partial z}{\partial P/\partial \rho},$$

$$\frac{d^2\rho}{dz^2} = \frac{2\frac{\partial P}{\partial \rho}\frac{\partial P}{\partial z}\frac{\partial^2 P}{\partial \rho \partial z} - \left(\frac{\partial P}{\partial \rho}\right)^2 \frac{\partial^2 P}{\partial z^2} - \left(\frac{\partial P}{\partial z}\right)^2 \frac{\partial^2 P}{\partial \rho^2}}{(\partial P/\partial \rho)^3}. \quad (48)$$

Thus the curvature at any point on the equiphase surface is

$$\kappa(\rho, z) = \frac{2\frac{\partial P}{\partial \rho}\frac{\partial P}{\partial z}\frac{\partial^2 P}{\partial \rho \partial z} - \left(\frac{\partial P}{\partial \rho}\right)^2 \frac{\partial^2 P}{\partial z^2} - \left(\frac{\partial P}{\partial z}\right)^2 \frac{\partial^2 P}{\partial \rho^2}}{[(\partial P/\partial \rho)^2 + (\partial P/\partial z)^2]^{3/2}}. \quad (49)$$

The radius of curvature is $\kappa^{-1}$. Near the beam axis we can write

$$P(\rho, z) = P(0, z) + \frac{1}{2}\rho^2 \left(\frac{\partial^2 P}{\partial \rho^2}\right)_{\rho=0} + O(\rho^4) \quad (50)$$

and the curvature becomes primarily a function of $z$:

$$\kappa(\rho, z) = -\left[\frac{\partial^2 P/\partial \rho^2}{\partial P/\partial z}\right]_{\rho=0} + O(\rho^2). \quad (51)$$

The on-axis curvature of the phase function $P$ of $\psi_{00}$, given in (44), is

$$\kappa_0(z) = \{z(b^2 + z^2)\cosh\beta \sinh\beta - 2z(b/k)$$
$$\times (\cosh^2\beta - \cos^2 kz) + (b^2 + z^2)\cos kz \sin kz\}$$
$$\times \{(b^2 + z^2)[(b^2 + z^2)\cosh\beta \sinh\beta$$
$$- (b/k)(\cosh^2\beta - \cos^2 kz)]\}^{-1}$$
$$= \frac{z}{b^2 + z^2}\frac{b^2 + z^2 - 2b/k}{b^2 + z^2 - b/k} + O(e^{-2\beta})$$
$$= \frac{(kz)^2 + kz \cos kz \sin kz - 2\sin^2 kz}{z[(kz)^2 - \sin^2 kz]} + O(\beta^2). \quad (52)$$

For comparison, the on-axis curvature of the equiphase surfaces of the Gaussian beam $\psi_G$ is $z/(b^2 + z^2 - b/k)$, which agrees with the curvature of the $\psi_{00}$ phase when $\beta = kb \gg 1$. We see from the last expression in (51) that $\kappa_0(z)$ has a functional form as $\beta \to 0$ which is well behaved in the focal region: $\kappa_0/k \to (\frac{2}{15})kz + O(kz)^3$. The curvature of the Gaussian approximate solution, in contrast, which can be written as $k^2 z/\lfloor \beta^2 + (kz)^2 - \beta \rfloor$, becomes infinite at $(kz)^2 = \beta(1 - \beta)$ and tends to $z^{-1}$ as $\beta \to 0$.

The curvature $\kappa_0(z)$ is shown in figure 8 for several values of $\beta$, including its limiting form as $\beta \to 0$. The $\beta = 0$ function has extrema at $kz = \pm\pi$, and has zero slope at $kz/\pi$ equal to positive or negative integers.

We now turn to the phase function $P_E$ of the transverse electric field components, given by (44) and (45). The on-axis curvature is more complicated than that of the phase function $P$, but is in agreement with it for large $\beta$:

$$\kappa_0^E(z) = \frac{z}{z^2 + b^2}\frac{\beta^4 - 6\beta^3 + 2(6 + \zeta^2)\beta^2 - 6(1 + \zeta^2)\beta + \zeta^4}{\beta^4 - 4\beta^3 + 2(3 + \zeta^2)\beta^2 - (3 + 4\zeta^2)\beta + \zeta^4}$$
$$+ O(e^{-2\beta}) \quad (53)$$

where $\zeta = kz$. The difference between the curvatures is greatest at small $\beta$: we find, again with $kz = \zeta$,

$$\kappa_0^E(z) = \{\zeta^4 - \zeta^3 \cos\zeta \sin\zeta - 6\zeta^2 \cos^2\zeta + 12\zeta \cos\zeta \sin\zeta$$
$$- 6\sin^2\zeta\}\{z[\zeta^4 - 2\zeta^2 \cos^2\zeta - \zeta^2$$
$$+ 6\zeta \cos\zeta \sin\zeta - 3\sin^2\zeta]\}^{-1} + O(\beta^2). \quad (54)$$

For small $z$ we find the $\beta = 0$ limit of $\kappa_0^E(z)$ tends to zero in the focal region as expected: $\kappa_0^E/k \to \frac{2}{35}kz + O(kz)^3$. In this limit the curvature is $\frac{3}{7}$ of the curvature of the equiphase surfaces of

497

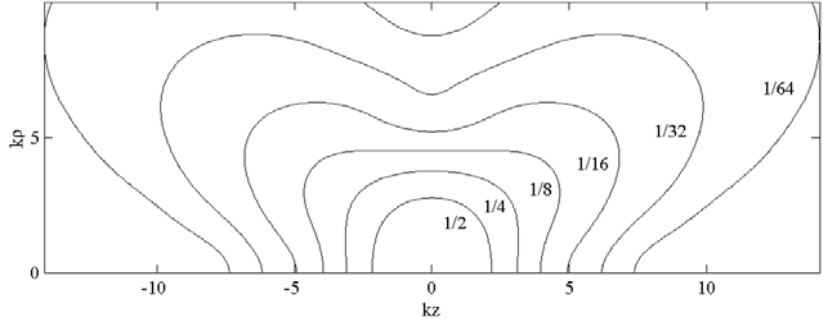

**Figure 10.** Energy density contours for the 'TEM' $\psi_{00}$ beam, drawn for $\beta = 2$. The energy density is normalized to unity at the origin. The three-dimensional surfaces of constant energy density are obtained by rotating the diagram about the $z$ axis.

$\psi$, for small $z$. Thus there can be a large difference between the curvatures of the equiphase surfaces of the electric field and of the vector potential.

The phase velocities on the beam axis are directed along the beam axis and have magnitude $v_p(z) = ck/(\partial P/\partial z)_{\rho=0}$. Figure 9 shows $v_p^E(z) = ck/(\partial P_E/\partial z)_{\rho=0}$ for several values of $\beta$. At the origin we have

$$\frac{v_p^E(0)}{c} = \frac{\beta(\beta^3 \cosh\beta \sinh\beta + 2\beta^2 \cosh^2\beta - 3\sinh^2\beta + \beta^2)}{(\beta^4 - 3\beta^2 + 9)\cosh^2\beta - (\beta^4 + 6\beta^2 + 9)}$$
$$= 1 + 2\beta^{-1} + 3\beta^{-2} + O(\beta^{-3}) = 5\beta^{-1} + \tfrac{1}{7}\beta + O(\beta^3). \quad (55)$$

At large $\beta$ the phase velocity of the electric field tends to $c$, for all values of $z$, but the phase speed at the origin is always larger than $c$ and increases without limit as $\beta$ decreases to zero.

For the 'TEM' beam we have [8, section 4]

$$\boldsymbol{B} = \frac{A_0}{k}\left(\frac{\partial^2\psi}{\partial x\,\partial z} + k\frac{\partial\psi}{\partial y}, \frac{\partial^2\psi}{\partial y\,\partial z} - k\frac{\partial\psi}{\partial x}, \frac{\partial^2\psi}{\partial z^2} + k^2\psi\right) \quad (56)$$

and $\boldsymbol{E} = i\boldsymbol{B}$. When $\psi$ is independent of the azimuthal angle,

$$\boldsymbol{B} = \frac{A_0}{k}\left(\cos\phi\frac{\partial^2\psi}{\partial\rho\,\partial z} + k\sin\phi\frac{\partial\psi}{\partial\rho},\right.$$
$$\left.\sin\phi\frac{\partial^2\psi}{\partial\rho\,\partial z} - k\cos\phi\frac{\partial\psi}{\partial\rho}, \frac{\partial^2\psi}{\partial z^2} + k^2\psi\right). \quad (57)$$

There are therefore three families of equiphase surfaces, one each for $B_x$, $B_y$ and $B_z$ (the equiphase surfaces of the components of $\boldsymbol{E}$ are related to those of $\boldsymbol{B}$ by a constant shift of $\pi/2$ in the phase values). Of these, the surfaces for $B_x$ and $B_y$ depend on the azimuthal angle; those for $B_z$ do not.

## 6. The 'TEM' energy flux and energy velocity

For the 'TEM' steady beam the complex fields are given by (55) and (56) with $\boldsymbol{E} = i\boldsymbol{B}$. The energy density $u$ and energy flux $\boldsymbol{S}$ are both independent of time. The steady beam expressions (37) for $u$ and $\boldsymbol{S}$ in terms of $\boldsymbol{B}_r$ and $\boldsymbol{B}_i$ can be rewritten as

$$u = \frac{1}{8\pi}\boldsymbol{B}\cdot\boldsymbol{B}^*, \qquad \boldsymbol{S} = \frac{ic}{8\pi}\boldsymbol{B}\times\boldsymbol{B}^*. \quad (58)$$

When $\psi$ is independent of the azimuthal angle $\phi$, we find from (56) that

$$u = \frac{A_0^2}{8\pi k^2}\left\{\left|\frac{\partial^2\psi}{\partial\rho\,\partial z}\right|^2 + k^2\left|\frac{\partial\psi}{\partial\rho}\right|^2 + \left|\frac{\partial^2\psi}{\partial z^2} + k^2\psi\right|^2\right\} \quad (59)$$

where $A_0$ is the (assumed real) amplitude of the vector potential:

$$\boldsymbol{A}_{TM} + i\boldsymbol{A}_{TE} = A_0\left(\frac{1}{k}\frac{\partial\psi}{\partial y}, -\frac{1}{k}\frac{\partial\psi}{\partial x}, \psi\right). \quad (60)$$

The energy flux vector obtained by expanding (57) is

$$\boldsymbol{S} = \frac{c}{4\pi}\operatorname{Im}(B_y^* B_z, B_z^* B_x, B_x^* B_y). \quad (61)$$

The $z$ component is independent of $\phi$:

$$S_z = \frac{c}{4\pi}\frac{A_0^2}{k}\operatorname{Im}\left\{\frac{\partial\psi^*}{\partial\rho}\frac{\partial^2\psi}{\partial\rho\,\partial z}\right\}. \quad (62)$$

The $x$ and $y$ components simplify on replacing $(\partial^2\psi/\partial z^2) + k^2\psi$ by $-(\partial^2\psi/\partial\rho^2) + (1/\rho)\partial\psi/\partial\rho$. We find that $S_x = S_\rho\cos\phi - S_\phi\sin\phi$ and $S_y = S_\rho\sin\phi + S_\phi\cos\phi$, where $S_\rho$ and $S_\phi$ are the radial and azimuthal components:

$$S_\rho = \frac{c}{4\pi}\frac{A_0^2}{k}\operatorname{Im}\left(\frac{\partial\psi^*}{\partial\rho}\frac{\partial^2\psi}{\partial\rho^2}\right),$$
$$S_\phi = \frac{c}{4\pi}\frac{A_0^2}{k^2}\operatorname{Im}\left\{\frac{\partial^2\psi^*}{\partial\rho\,\partial z}\left(\frac{\partial^2\psi}{\partial\rho^2} + \frac{1}{\rho}\frac{\partial\psi}{\partial\rho}\right)\right\}. \quad (63)$$

The magnitude of the transverse component of $\boldsymbol{S}$ is independent of $\phi$: $S_x^2 + S_y^2 = S_\rho^2 + S_\phi^2$. The azimuthal component $S_\phi$ contributes to the angular momentum of the beam: $\boldsymbol{p} = \boldsymbol{S}/c^2$ is the momentum density, so $\boldsymbol{r}\times\boldsymbol{p}$ is the angular momentum density [16], which has the $z$ component

$$(\boldsymbol{r}\times\boldsymbol{p})_z = xp_y - yp_x = c^{-2}\rho S_\phi. \quad (64)$$

The 'TEM' beam energy density for the $\psi_{00}$ wavefunction is shown in figure 10 and the energy flux is shown in figure 11. We see that the energy density is non-zero on the beam axis $\rho = 0$, whereas $\boldsymbol{E}\times\boldsymbol{B}$ is zero on the axis. This is because $\boldsymbol{B}$ and $\boldsymbol{E}$ both have only longitudinal components on the beam axis, since both $B_x$ and $B_y$ are zero there (see (56) and (60)). Thus the beam is hollow in energy flux and momentum density, and the energy velocity $\boldsymbol{v}_e = \boldsymbol{S}/u$ is *zero* on the beam axis: there is energy on the axis, but it is not moving. Figure 12 shows the energy velocity field of the $\psi_{00}$ 'TEM' beam.

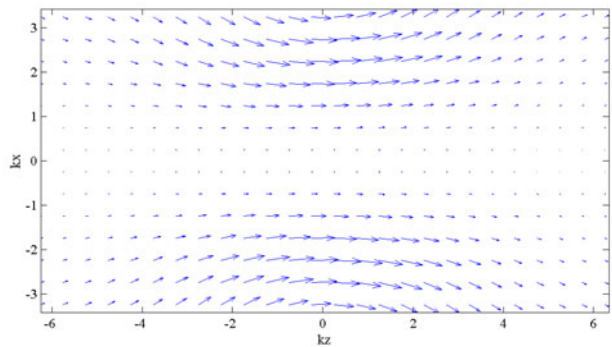

**Figure 11.** Energy flux ($S_z$, $S_\rho$) for the 'TEM' $\psi_{00}$ beam, drawn for $\beta = 2$. The azimuthal component $S_\phi$ is not shown. Note that the beam is hollow in energy flux and in momentum.

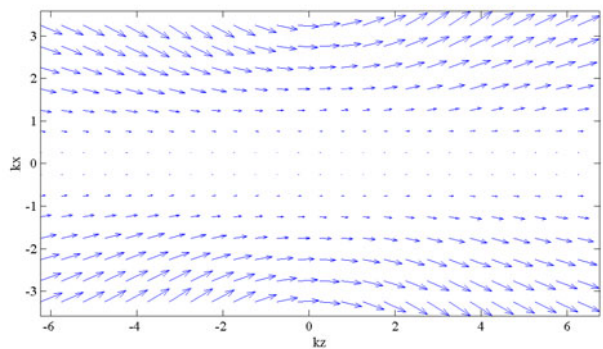

**Figure 12.** Energy velocity field for the 'TEM' $\psi_{00}$ beam, drawn for $\beta = 2$. The azimuthal component (not shown) is also zero on the beam axis.

## 7. Discussion

We have seen that the phase and particle transport velocities for focused beams can differ markedly from the broad-beam values $\hbar k/2m$ and $\hbar k/m$, respectively. For example, when $\beta = 2$, at the centre of the focal plane of the $\psi_{00}$ beam the phase velocity is $1.86(\hbar k/2m)$ and the particle transport velocity is $(\hbar k/m)/1.86$. For the $\psi_G$ and $\psi_{10}$ beams the multiplier 1.86 is replaced by 2 and 1.16, respectively. Recent developments in helium atom beam focusing using Fresnel zone plates [17] may provide the techniques to test these predictions.

In electromagnetic beams the energy velocity never exceeds the speed of light. It can be zero, as we saw in the case of the 'TEM' beam where the energy flux is zero on the beam axis, but the energy density is not. The phase velocities can be larger than the speed of light: this corresponds to the effective wavelength $2\pi/|\nabla P|$ being larger than $2\pi/k$ (i.e. $|\nabla P| < k$). The phase velocities, one each for the components of $E$ and $B$, where these components differ in the form of the phase function $P(r)$, are independent of time. The energy velocity in general oscillates about its mean value at angular frequency $2\omega$, except in the case of *steady beams*, for which the energy flux, energy density and energy velocity are all independent of time. It is interesting that the phase velocities of the various components of the electric and magnetic fields can differ from each other in direction and in magnitude. Up to six different sets of equiphase surfaces can exist in an electromagnetic beam. As a function of time, up to six sets of wavefronts are propagating, at various speeds and in various directions. Since the curvature of the mirrors bounding resonant laser cavities is matched with the curvature of the wavefront, one should know which wavefront is the relevant one. Presumably it is that of the transverse electric field components, but the literature appears to be silent on this question. As we saw in section 5, the curvatures of the various wavefronts can be very different when $kb$ and $kz$ are both small.

In conclusion, we remark on the perhaps surprising result that orthodox wave optics leads us to predict that electromagnetic energy can travel in free space at less than the speed of light, and that it can even stand still in parts of some propagating beams.

## Acknowledgments

The author is grateful to Paul Callaghan, Thomas Iorns and Damien Martin and to an anonymous referee for stimulating questions and comments.

## References

[1] Jackson J D 1975 *Classical Electrodynamics* 2nd edn (New York: Wiley)
[2] Boyd G D and Gordon J P 1961 Confocal multimode resonator for millimeter through optical wavelength masers *Bell Syst. Tech. J.* **40** 489–508
[3] Kogelnik H and Li T 1966 Laser beams and resonators *Appl. Opt.* **5** 1550–67
[4] Lax M, Louisell W H and McKnight W B 1975 From Maxwell to paraxial wave optics *Phys. Rev.* A **11** 1365–70
[5] Davis L W 1979 Theory of electromagnetic beams *Phys. Rev.* A **19** 1177–9
[6] Sheppard C J R and Saghati S 1998 Beam modes beyond the paraxial approximation: a scalar treatment *Phys. Rev.* A **57** 2971–9
[7] Ulanowski Z and Ludlow I K 2000 Scalar field of nonparaxial Gaussian beams *Opt. Lett.* **25** 1792–4
[8] Lekner J 2001 TM, TE and 'TEM' beam modes: exact solutions and their problems *J. Opt. A: Pure Appl. Opt.* **3** 407–12
[9] Brillouin L 1946 *Wave Propagation in Periodic Structures* (New York: McGraw-Hill) (reprinted by Dover 1953)
[10] Born M and Wolf E 1999 *Principles of Optics* 7th edn (Cambridge: Cambridge University Press)
[11] Merzbacher E 1998 *Quantum Mechanics* 3rd edn (New York: Wiley)
[12] Landesman B T and Barrett H H 1988 Gaussian amplitude functions that are exact solutions to the scalar Helmholtz equation *J. Opt. Soc. Am.* A **5** 1610–19
[13] Haus H A 1984 *Waves and Fields in Optoelectronics* (New York: Prentice-Hall)
[14] Siegman A E 1986 *Lasers* (Sausalito: University Science Books)
[15] Lekner J 2000 Multiple principal angles for a homogeneous layer *J. Opt. A: Pure Appl. Opt.* **2** 239–45
[16] Allen L, Padgett M J and Babiker M 1999 The orbital angular momentum of light *Prog. Opt.* **39** 291–372
[17] Doak R B, Grisenti R E, Rehbein S, Schmahl G, Toennies J P and Wöll Ch 1999 Toward realization of an atomic de Broglie microscope: helium atom focusing using Fresnel zone plates *Phys. Rev. Lett.* **83** 4229–32